\documentclass[%
 amsmath,amssymb,
 aps,
]{revtex4-1}

\usepackage{graphicx}
\usepackage{dcolumn}
\usepackage{bm}

\begin{document}

\title{Black Hole Membrane Paradigm from Boundary Scalar Field}
\author{Jingbo Wang}
\email{ shuijing@mail.bnu.edu.cn}
\affiliation{Department of Physics and Electronic Engineering, Hebei Normal University for Nationalities, Chengde, 067000, China}
 \date{\today}
\begin{abstract}
Black hole membrane paradigm suggests to consider the black hole horizon as a fluid membrane. The membrane has a particular energy-momentum tensor which characterizes the interactions with the falling matter. In this paper, we show that we can construct an action from the scalar field on the horizon which can give the same energy-momentum tensor for the membrane. That is, the membrane can be described effectively by the scalar field on it.
\end{abstract}
\pacs{04.70.Dy,04.60.Pp}
 \keywords{Membrane paradigm; Boundary scalar field; Energy--momentum tensor}
\bibliographystyle{unsrt}
\maketitle
\section{Introduction}
The ``membrane paradigm" \cite{mb0,mb1,mb2} suggests that to an outside observer, when interacts with the falling matters, the black hole behaves like a dynamical fluid membrane. This membrane is regarded as made from a 2-dimensional viscous fluid that is electrical charged and electrically conducting. It also has finite entropy and temperature, but can't conduct heat.

In recent years, the membrane paradigm in Einstein gravity was revisited in the emergent gravity paradigm \cite{emerg1,emerg3}. Such an approach highlighted the deeper connection between the membrane paradigm and the horizon thermodynamics. One connection is pointed in \cite{emerg1} relating the membrane surface pressure $p$ to the entropy $S$ of the horizon through an state equation
\begin{equation}\label{0}
  p A= S T,
\end{equation}
where $T$ is the Hawking temperature and $A$ is the area of the horizon.

In the previous works \cite{whcft1,wangbms4}, the boundary modes on the horizon of the BTZ black hole and the Kerr black hole were analyzed. They both have boundary degrees of freedom which can be described by a scalar field theory. From this scalar field we give the microstates for Ba$\tilde{n}$ados-Teitelboim-Zanelli (BTZ) black holes and Kerr black holes. Those microstates can account for the Bekenstein-Hawking entropy. We also show that those states can also give the Hawking radiation \cite{wanghr1}. Actually the Hawking radiation is the mixture of thermal radiation of right/left-moving sectors of this scalar field at different temperatures. Based on this result \cite{wanghr2}, for statics BTZ black holes and Schwarzschild black holes, we propose a simple solution for the information loss paradox. That is, the Hawking radiation is pure due to the entanglement between the left-moving sector and right-moving sector of the Hawking radiation. So the scalar field can describe the equilibrium state of the black hole physics. Since the membrane paradigm summarize the dynamics of the black hole, we want to know if the scalar field can also gives those properties of the membrane. In this paper, we will show that the answer is yes. That is, from the boundary scalar field, one can construct the membrane paradigm of the black hole. The energy-momentum tensor from the action gives the desired results.

The paper is organized as follows. In section II, we construct the action of the scalar field step by step. In section III, we calculate the parameters in the action for BTZ black holes and Kerr black holes. Section IV is the conclusion.
\section{Energy-momentum tensor from scalar field}
From hydrodynamics, it is well known that the energy-momentum tensor for a viscous fluid on $D$-dimensional spacetime takes the form
\begin{equation}\label{1}
  T_{AB}=\rho u_A u_B+p (g_{AB}+u_A u_B)+\pi_{AB},
\end{equation}
where $\rho$ is the energy density, $p$ the pressure, $u_A$ the 4-velocity and \cite{nas1}
\begin{equation}\label{2}\begin{split}
  \pi_{AB}=-2 \eta \sigma_{AB}-\zeta \theta \gamma_{AB},\\
  \sigma_{AB}=\frac{1}{2}(\nabla_A u_B+\nabla_B u_A+ a_A u_B+a_B u_A)-\frac{1}{D-1}\theta (g_{AB}+u_A u_B),\\
  \theta=\nabla^A u_A,\quad a_A=u_B \nabla^B u_A.
\end{split}\end{equation}

In membrane paradigm for black holes in $D+1$-dimensional sapcetime, the energy-momentum tensor projected onto a $D-1$-dimensional cross-section of the horizon takes the form of a viscous fluid \cite{mb3}
\begin{equation}\label{2a}
  T_{AB}=p \gamma_{AB}-2 \eta \sigma_{AB}-\zeta \theta \gamma_{AB},
\end{equation}
where $p=\frac{\kappa}{8\pi G}$ is the pressure, $\eta=\frac{1}{16\pi G}$ the shear viscosity, $\zeta=-\frac{D-2}{8\pi G (D-1)}=-\frac{2(D-2)}{D-1}\eta$ the bulk viscosity of the membrane and $\gamma_{AB}=g_{AB}+u_A u_B$. Compared with (\ref{2}), we can get that the energy density for the membrane is $\rho=0$. In the previous works, we use the boundary scalar field to describe the boundary degrees of freedom on the horizon. In the following we will show that the above energy-momentum tensor can also be obtained from the scalar field.

Firstly, we consider a free massless scalar field with the action \cite{wangbms4}
\begin{equation}\label{3}
  S_1=m_0 \int d^D x \sqrt{-g} [-\frac{1}{2} g^{AB}\partial_A \phi \partial_B \phi],
\end{equation}
where $m_0$ is a parameter determined by the black holes.

The energy-momentum tensor from this action is given by
\begin{equation}\label{3a}
  T_{AB}=-\frac{2}{\sqrt{-g}}\frac{\delta S_1}{\delta g^{AB}}=m_0 (\partial_A \phi \partial_B \phi-g_{AB}g^{CD}\partial_C \phi \partial_D \phi).
\end{equation}

Next we define the 4-velocity of the fluid flow described by the scalar field as
\begin{equation}\label{4}
  u_A=\frac{\partial_A \phi}{\sqrt{-g^{CD}\partial_C \phi \partial_D \phi}},
\end{equation}
which satisfy $g^{AB} u_A u_B=-1$.

The energy-momentum tensor for this free scalar field is given by
\begin{equation}\label{5}
  T_{AB}=m_0 (u_A u_B+g_{AB})(-g^{CD}\partial_C \phi \partial_D \phi),
\end{equation}
which gives
\begin{equation}\label{6}
  \rho=p=m_0 (-g^{CD}\partial_C \phi \partial_D \phi).
\end{equation}

To get the zero energy density, the simplest way is to add a cosmological constant term $\Lambda$ in the action, and gives
\begin{equation}\label{7}\begin{split}
  \rho=m_0 (-g^{CD}\partial_C \phi \partial_D \phi)-\Lambda=0 \Rightarrow \Lambda=m_0 (-g^{CD}\partial_C \phi \partial_D \phi).\\
  p=m_0 (-g^{CD}\partial_C \phi \partial_D \phi)+\Lambda=-m_0 g^{CD}\partial_C \phi \partial_D \phi.
\end{split}\end{equation}
Now we get the first terms of (\ref{2a}) for black holes. Next we will add more terms in the action to get the left terms in (\ref{2a}). After some trials, one can show that the action
\begin{equation}\label{8}
   S_2=\alpha_D \int d^D x (-g) (-g^{AB}\partial_A \phi \partial_B \phi)^{3/2} \nabla^C u_C=\alpha_D \int d^D x (-g) (-g^{AB}\partial_A \phi \partial_B \phi)^{3/2} \theta
\end{equation}
will gives the the last two terms in (\ref{2a}) with (see the appendix)
\begin{equation}\label{9}
  \eta=\alpha_D \sqrt{-g} (-g^{AB}\partial_A \phi \partial_B \phi)^{3/2},\quad \zeta=-\frac{2(D-2)}{D-1}\eta.
\end{equation}
So if we choose the suitable parameter $\alpha_D=\frac{1}{16\pi G \sqrt{-g}} (-g^{AB}\partial_A \phi \partial_B \phi)^{-3/2}$, we can get the correct shear and bulk viscosity of the membrane.

The full action for the scalar field on the membrane is
\begin{equation}\label{10}
  S=m_0 \int d^D x \sqrt{-g} (-\frac{1}{2} g^{AB}\partial_A \phi \partial_B \phi)+\int d^D x \sqrt{-g}\Lambda+\alpha_D \int d^D x (-g) (-g^{AB}\partial_A \phi \partial_B \phi)^{3/2} \nabla^C u_C,
\end{equation}
where $\Lambda=m_0 (-g^{CD}\partial_C \phi \partial_D \phi)$ and $\alpha_D=\frac{1}{16\pi G \sqrt{-g}} (-g^{AB}\partial_A \phi \partial_B \phi)^{-3/2}$.

The field equation for the scalar field is easy to get
\begin{equation}\label{10a}
  m_0 \nabla_a \nabla^a \phi-2\sqrt{-g} \alpha_D [(\nabla_b \nabla^b \phi)^2-\nabla_c \nabla_d \phi \nabla^c \nabla^d \phi]=0,
\end{equation}
which is a non-linear field equation.
\section{For BTZ black holes and Kerr black holes}
In this section we calculate those parameters for BTZ black holes and Kerr black holes.
\subsection{BTZ black hole}
The metric of the BTZ black hole is \cite{btz1}
\begin{equation}\label{1}
    ds^2=-N^2 dv^2+2 dv dr+r^2 (d\varphi+N^\varphi dv)^2,
\end{equation}
where $N^2=-8 M+\frac{r^2}{L^2}+\frac{16 J^2}{r^2}, N^\varphi=-\frac{4 J}{r^2}$, $L$ the radius of the AdS spacetime and $(M,J)$ are mass and angular momentum of the black hole respectively. The black hole has the event horizon at $r=r_+$.

For BTZ black hole, the boundary scalar field is \cite{wangbms4}
\begin{equation}\label{11}\begin{split}
  \phi(v',\varphi)=\phi_0+p_v v'+p_\varphi \varphi+ \sqrt{\frac{1}{m_0 A}}\sum_{n\neq 0}\sqrt{\frac{1}{2 \omega_n}}[a_n e^{-i(\omega_n v'-k_n \varphi)}+a^+_n e^{i(\omega_n v'-k_n \varphi)}],
  \end{split}\end{equation}
where $v'=\frac{v}{\gamma}=\frac{r_+}{L}v,\omega_n=\frac{|n|}{r_+},k_n=n,m_0=\frac{L}{8 \pi}$ and $A=2\pi r_+$ is the length of the circle. The zero-mode part is given by $ p_v= -\frac{1}{L},\quad p_\varphi=\frac{r_- }{L}$.

The effective metric is
\begin{equation}\label{12}
  \tilde{ds}^2=-dv'^2+r_+^2 d\varphi^2,
\end{equation}
so one can get
\begin{equation}\label{13}
-g^{CD}\partial_C \phi \partial_D \phi=p_v^2-p_\varphi^2/r_+^2= \frac{r_+^2-r_-^2}{r_+^2 L^2}=\frac{2\pi}{r_+} T_H.
\end{equation}
The pressure of the membrane observed at infinity is
\begin{equation}\label{14}
  p=m_0(-g^{CD}\partial_C \phi \partial_D \phi)/\gamma= \frac{L}{8 \pi G} \frac{2\pi}{r_+} T_H \frac{r_+}{L}=\frac{T_H}{4 G}=\frac{\kappa}{8\pi G},
\end{equation}
which is just the pressure for BTZ black hole and satisfy the state equation (\ref{0}). The factor $\gamma$ appears because at infinity the pressure is associated with time parameter $v=\gamma v'$.

The parameters in the action (\ref{10}) for BTZ black hole are
\begin{equation}\label{15}
 \Lambda=\frac{L}{4 G r_+} T_H,\quad  \alpha_D=\frac{1}{16\pi G r_+}  (\frac{2\pi}{r_+} T_H)^{-3/2}.
\end{equation}
\subsection{Kerr black hole}
The metric of the Kerr black hole can be written as \cite{kerr1}
\begin{equation}\label{26}
  ds^2=-(1-\frac{2 M r}{\rho^2})dv^2+2 dv dr-2 a \sin^2 \theta dr d\varphi-\frac{4 a M r \sin^2 \theta}{\rho^2}dv d\varphi+\rho^2 d\theta^2+\frac{\Sigma^2 \sin^2 \theta}{\rho^2}d\varphi^2,
\end{equation}
where $\rho^2=r^2+a^2 \cos^2 \theta, \Delta^2=r^2-2 M r+a^2, \Sigma^2=(r^2+a^2)\rho^2+2 a^2 M r \sin^2 \theta$ with $(M, J=M a)$ the mass and angular momentum of the Kerr black hole. The event horizon is localized at $r=r_+=M+\sqrt{M^2-a^2}$.

For Kerr black hole, the boundary scalar field is \cite{wangbms4}
\begin{equation}\label{16}\begin{split}
  \phi(v',\theta,\varphi)=\phi_0+p_v v'+p_\varphi\varphi+\sqrt{\frac{1}{m_0 A}}\sum_{l\neq 0}\sum_{m=-l}^{m=l}\sqrt{\frac{1}{2\omega_l}}[a_{l,m} e^{-i \omega_l v'}Y^m_l(\theta,\varphi)+a^+_{l,m} e^{i \omega_l v'}(Y^m_l)^*(\theta,\varphi)],\\
  \end{split}\end{equation}
where $v'=\frac{v}{\gamma}=\frac{r_+^2}{r_+^2+a^2}v,\omega^2_l=\frac{l(l+1)}{r_+^2},m_0=\frac{M}{2\pi}$, $Y^m_l(\theta,\varphi)$ are spherical harmonics and $A=4\pi r_+^2$. The zero-mode part is given by $ p_v= -\frac{r_+^2+a^2}{4 M r_+^2},\quad p_\varphi=\frac{a (r_+^2+a^2)}{4 M r_+^2}$.

The effective metric is
\begin{equation}\label{17}
  \tilde{ds}^2=-dv'^2+r^2_+(d\theta^2+\sin^2 \theta d\varphi^2),
\end{equation}
so one can get
\begin{equation}\label{18}
-g^{CD}\partial_C \phi \partial_D \phi=p_v^2-|\frac{p_\varphi^2}{r_+^2 \sin^2 \theta}|= \frac{r_+^2-a^2}{4 r_+^2}=\frac{2\pi M}{r^2_+} T_H.
\end{equation}
The pressure of the membrane observed at infinity is
\begin{equation}\label{19}
  p=m_0(-g^{CD}\partial_C \phi \partial_D \phi)/\gamma= \frac{M}{2 \pi} \frac{2\pi M}{r^2_+} T_H \frac{r^2_+}{r_0^2}=\frac{T_H}{4 G} \frac{2 M G}{r_+}=\frac{T_H}{4 G} \frac{A_0}{A}.
\end{equation}
The term $\frac{A_0}{A}$ appears because the horizon of Kerr black hole has radius $r_+$ but the area $A_0=4\pi(r_+^2+a^2)$. It is easy to show that the pressure satisfies the state equation $pA=TS$.

The parameters in the action (\ref{10}) for the Kerr black hole are
\begin{equation}\label{20}
 \Lambda=\frac{M^2}{r^2_+} T_H,\quad  \alpha_D=\frac{1}{16\pi G r^2_+}  (\frac{2\pi M}{r^2_+} T_H)^{-3/2}.
\end{equation}
\section{Conclusion}
In this paper, we get the energy-momentum tensor for the membrane paradigm from the scalar field on the horizon. That is, from the action (\ref{10}) one calculate the energy-momentum tensor and get the correct result (\ref{2a}). The parameters in the action are determined by the black holes. The action contains three terms: the first one is just the free scalar field, and the second one is a cosmological constant term and the third one is a term proportional to the expansion $\theta$. The third term contain self-interaction, so the final theory is an interacting scalar field theory.

The membrane paradigm describe the interactions of black hole with the in-falling matters. So the boundary scalar field not only can describe the equilibrium state of the black holes, but also the dynamics of the black hole. Quantized this scalar theory can also give a microscopic theory that describe the membrane \cite{witten1}.
\acknowledgments
The author would like to thank Prof.Wei Han for many helps.
\section{Appendix}
In this appendix, we will show that the action (\ref{8}) gives the result (\ref{9}). The action can be divided into three parts,
\begin{equation}\label{a1}
   S'=\alpha_D \int d^D x \underbrace{(-g)}\underbrace{(-g^{AB}\partial_A \phi \partial_B \phi)} \underbrace{(-g^{AB}\partial_A \phi \partial_B \phi)^{1/2} \theta}.
\end{equation}
The energy-momentum tensor is given by (\ref{3a}). We denote $X=(-g^{AB}\partial_A \phi \partial_B \phi)$.

Due to
\begin{equation}\label{a2}
 \frac{\delta (-g)}{\delta g^{AB}}=-(-g) g_{AB},
\end{equation}
the first part gives
\begin{equation}\label{a3}
  T_{1AB}=2\sqrt{-g} g_{AB} X^{3/2} \theta.
\end{equation}

The second part gives
\begin{equation}\label{a4}
  T_{2AB}=2\sqrt{-g} \partial_A \phi \partial_B \phi X^{1/2} \theta=2\sqrt{-g} u_A u_B X^{3/2} \theta.
\end{equation}

The third part is a little complicate, and it was shown that \cite{nas1}
\begin{equation}\label{a5}
 X^{1/2} \theta=\nabla_a \nabla^a \phi+ X^{-1}\nabla_c \phi \nabla_d \phi \nabla^c \nabla^d \phi,
\end{equation}
and the variation under $g^{ab}$ gives
\begin{equation}\label{a6}
 X^{1/2} \frac{1}{2}(\nabla_A u_B+\nabla_B u_A+ a_A u_B+a_B u_A).
\end{equation}
So the third part gives
\begin{equation}\label{a7}
  T_{3AB}=2\sqrt{-g} X X^{1/2} \frac{1}{2}(\nabla_A u_B+\nabla_B u_A+ a_A u_B+a_B u_A).
\end{equation}

The total energy-momentum tensor is
\begin{equation}\label{a8}
  T_{AB}=T_{1AB}+T_{2AB}+T_{3AB}=2\sqrt{-g} X^{3/2} [ \frac{1}{2}(\nabla_A u_B+\nabla_B u_A+ a_A u_B+a_B u_A)+\theta (g_{AB}+u_A u_B)].
\end{equation}
Compared with the standard expression (\ref{2}), it is easy to show that one can get the result (\ref{9}).

\end{document}